# Proposal of an optical Bell's experiment to test the boundary between determinism and indeterminism in Quantum Mechanics.

Alejandro Hnilo, Marcelo Kovalsky, Mónica Agüero and Myriam Nonaka.
*CEILAP, Centro de Investigaciones en Láseres y Aplicaciones, UNIDEF (MINDEF-CONICET);*
*CITEDEF, J.B. de La Salle 4397, (1603) Villa Martelli, Argentina.*
*email: ahnilo@citedef.gob.ar*
August 20th, 2024.

It was recently noted the existence of an apparently discontinuous boundary between determinism and indeterminism in Quantum Mechanics. We propose to explore this boundary in an optical Bell's experiment by recording the distribution, of the number of strings of outcomes of a given parity interrupted by outcomes of the other parity, as a function of their length. The features of these distributions for small rotations of the angle settings near critical points may indicate whether the underlying process is in-deterministic or not. Therefore, they may show that the boundary is discontinuous, or else, that determinism decays smoothly. The conditions the experimental setup must fulfill are discussed.

## 1. Introduction.

The indeterminism intrinsic in the predictions of Quantum Mechanics (QM) has faced controversy since its very inception. The Einstein-Podolsky-Rosen argument, in particular, deals with the case of strict correlations between the outcomes of observations performed on distant entangled particles, as an example of a case of deterministic behavior in contradiction with that intrinsic indeterminism.

K.Svozil has recently noted and discussed an intriguing situation which involves an apparently discontinuous boundary between determinism and indeterminism in QM [1]. His original reasoning involves the state $|\Psi\rangle = (1/\sqrt{2}).(|0\rangle+|1\rangle)$ and projections over the observable $|\Psi\rangle\langle\Psi|$, the observable orthogonal to this one, and to "tiny" rotations of both. In order to deal with the same state and observables that are later used in the proposed experiment, let consider instead the case of the fully symmetrical Bell state entangled in polarization $|\Phi^+\rangle = (1/\sqrt{2}).(|x_A,x_B\rangle + |y_A,y_B\rangle)$ in the usual notation, and polarization observables. The situation is essentially equivalent. For state $|\Phi^+\rangle$, the probabilities of observing coincidences are:

$$P^{++} = P^{--} = \tfrac{1}{2}.cos^2(\alpha-\beta); \ P^{-+} = P^{+-} = \tfrac{1}{2}.sin^2(\alpha-\beta) \quad (1)$$

where "+" ("-") means detection in the "transmitted" ("reflected") port of the polarizer in the corresponding station (A or B), and α (β) is the angle orientation of the polarizer in station A (B). At the critical point α=β, all coincidences have "even" parity: ++ or --. In this case, the system is deterministic: knowing the outcome in his own station, A (B) knows with certainty the outcome B (A) has observed, regardless the distance or timing. This certainty is the reason why a shared key can be built in Quantum Key Distribution (QKD). Besides, in the case α=β the probabilities in Eqs.1 can be derived from a simple classical, deterministic and *local* model (also for $|\alpha-\beta| = 45^o$, $90^o$) [2]. The key in QKD is useful anyway, because an eavesdropper ignores, in practice, if the outcome is ++ or --, but the arguments supporting "certified quantum randomness" do not apply to these cases. The key is, at best, random by ignorance. This fundamental vulnerability of QKD is rarely mentioned.

If α=β, as stated in [1]: *…there is no uncertainty, and no possibility to obtain randomness. Randomness comes about if "detuned experiments" are performed… any "tiny" rotation 0≠φ<<1…suffices to yield irreducible randomness.* Consider then a "tiny" rotation of the setting, so that α≠β with $|\alpha-\beta| << 1$. Svozil argues that, by applying a randomness extractor, irreducible randomness can be then achieved. *This sudden, discontinuous change from determinism into complete indeterminism by some "smooth, continuous" manipulation boggles a mind prepared to "evangelically" accept the quantum canon.* This is hence an intriguing situation.

In this paper, we study the possibility to explore this apparently discontinuous change experimentally. We show that in an experiment with high, but attainable, levels of entanglement and efficiency, there is a window (for the values of $|\alpha-\beta|$) within which deterministic features can be identified, if they exist. If determinism were actually identified within that window, it would be demonstrated that it does not disappear abruptly. Besides, that observation would support the idea that there is no intrinsic indeterminism in QM, but only indeterminism originated in the environment (which includes the classical instruments of observation), as it is argued in [1].

## 2. Time series of outcomes. Ideal case.

Consider a series of $m >>1$ coincident outcomes in an ideal Bell's setup with state $|\Phi^+\rangle$. If α=β, all elements in the series are "even" (i.e., they are all ++ or --). As soon as α≠β, some "odd" elements (+- or -+) start to appear in the series. If $|\alpha-\beta| << 1$ they are few. Let say that an odd element interrupts a "string" of even elements. If the events "coincidence observed" are independent and identically distributed random, then the number of strings (of uninterrupted even results) of length $k$ is:

$$n(k) = m.\varphi^2.(1-\varphi)^k \quad (2)$$

where $m$ is assumed sufficiently large so that the longest string is much shorter than the series (see below), and $\varphi$ is the probability of observing an odd element:

$$\varphi = sin^2(\alpha-\beta) \qquad (3)$$

Eq.2 is derived from complete indeterminism, and is also the QM prediction. The total number of strings is:

$$N_{strings} \approx m \cdot \varphi \qquad (4)$$

The largest meaningful value of $k$ is derived from $n(k_{max}) = 1$ and reads:

$$k_{max} \approx -ln(m \cdot \varphi^2)/ ln(1-\varphi) \qquad (5)$$

Eq.4 can be derived directly from Eqs.1, because there is one string for each odd element. Eq.2 is valid if $m >> k_{max}$; using Eq.4 and 5 and that $\varphi<<1$, then Eq.2 is valid as far as $N_{strings} >> ln(\varphi \cdot N_{strings})$, what always holds.

As said, the case $\varphi=0$ is deterministic. In this case there is only one string of length $m$ (the series itself). Let suppose that determinism does not vanish abruptly, but that it extends smoothly into the $\varphi \neq 0$ region. Because of the continuous change, it is intuitive that the series should decompose in a set of long strings as $\varphi$ moves away from $\varphi=0$. In order to provide support and definite numbers to this intuition, we refer to numerical simulations of a simple deterministic and *nonlocal* hidden variables model [3], let call it WQM here. It is briefly described in the Appendix; at this point, it suffices to say that WQM accurately reproduces the probabilities in Eq.1. Hence it accurately reproduces the result in Eq.4 too. Yet, as WQM is deterministic, the strings' distribution it predicts is, in general, different from $n(k)$, see Figure 1.

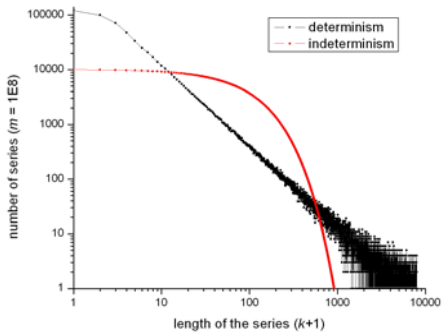

Figure 1: Numerically calculated distribution of strings' lengths in the ideal case, according to the nonlocal deterministic model WQM (black curve) and to in-deterministic $n(k)$ (red curve); $\varphi=0.01$ ($|\alpha-\beta| \approx 5.7°$), $m=10^8$.

As intuitively expected, the deterministic model predicts an excess of long strings (well above the value predicted by Eq.4) for $\varphi<<1$, which should be noticeable in an experiment. The two distributions are visibly different, yet their average values are close. The difference between the two distributions vanishes as $\varphi$ increases, see Figure 2. For $\varphi=0.1$ ($|\alpha-\beta|=18.4°$) the number of short strings in both distributions are nearly equal. For $\varphi>0.1$ short strings are more numerous in $n(k)$ than in WQM. The number of long strings instead, remains smaller in $n(k)$ than in WQM for the full range of $\varphi$, as intuitively expected. Yet, for $\varphi=0.5$ ($|\alpha-\beta|=45°$) the two distributions are indistinguishable.

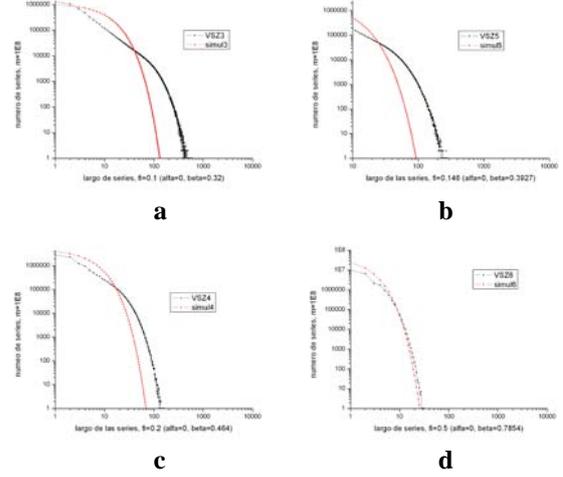

Figure 2: Numerically calculated distributions of strings' lengths in the ideal case for increasing values of $\varphi$: (a) $\varphi=0.1$, (b) $\varphi=0.146$, (c) $\varphi=0.2$, (d) $\varphi=0.5$. Other data as in Fig.1.

If $|\alpha-\beta|>45°$ it is appropriate switching the definition of "string", which now becomes a sequence of many odd elements separated by an even one. The curves then change symmetrically to Figs.2 until $|\alpha-\beta| = 90°$, where full determinism is reached again (i.e., all elements in the series are odd). Assuming that $n(k)$ *defines* indeterminism, Figs.2 suggest there can be a continuous, smooth observable variation from full determinism in the cases $|\alpha-\beta| \approx 0°$, $90°$ (where the distributions are very different, as in Fig.1) to full indeterminism in the case $|\alpha-\beta| = 45°$ (where the distributions are indistinguishable, Fig.2d).

In Figs.1-2, determinism and indeterminism can be distinguished over a broad range of values of $\varphi$. But these curves are plotted for the ideal case. Let see next the conditions experiments must fulfill to distinguish them in practice.

### 3. Time series of outcomes. Real case.
*3.1 Noise (dark counts).*

Single photon detectors emit signals even if they are in complete darkness. This effect is known as "dark counts", and is (supposedly) in-deterministic. Although dark counts occur uncorrelated in each detector in a Bell's experiment, they anyway produce a finite number of accidental coincidences with rate $(R_{dark})^2 \cdot T_w$, where $R_{dark}$ is the rate of dark counts (which is a detector's feature) and $T_w$ is the chosen time-coincidence window. Half these accidental coincidences are odd events. They interrupt otherwise long strings and may make the deterministic regime undetectable. When measuring entanglement (say, the $S_{CHSH}$ parameter [4]), the total number of accidental coincidences can be estimated and is often subtracted. In the case of interest here, it is impossible to get rid of accidental coincidences. For, there is no way to know which odd elements in the time series are "true" (i.e, a consequence of the tiny rotation), and which ones are caused by accidental coincidences. The only possible approach is to restrict the study to the parameters' space where the rate of "true" strings $\varphi.R_{coinc}$ is much larger than the estimated rate of strings produced by the accidental coincidences, that is:

$$\varphi.R_{coinc} >> \tfrac{1}{2}.(R_{dark})^2.T_w \qquad (6)$$

This is the limit an experiment can get close to the boundary at $\varphi=0$. The rate of true coincidences $R_{coinc}$ can be calculated by subtracting the estimated rate of accidental coincidences from the observed total rate of coincidences. For values of $\varphi$ so small that Eq.6 is not valid, dark counts make the observable distribution in-deterministic, even if the determinism present at $\varphi=0$ still exists. In order the hypothesized smooth decay of determinism to be observable, it must exist at values of $\varphi$ large enough such that Eq.6 is valid.

Typical numbers for an experiment with standard single photon detectors (silicon avalanche photodiodes) are $R_{dark} = 100$ s$^{-1}$, $R_{coinc} = 5\times10^4$ s$^{-1}$, $T_w = 10$ ns then $\varphi >> 10^{-9}$, or $|\alpha-\beta| >> 0.03$ mrad. This is a very small angle, so that Eq.6 does not impose a serious limitation in practice. Be aware that for angles below that value the observed distribution is surely in-deterministic, but that this does *not* imply intrinsic quantum indeterminism to exist.

*3.2 Imperfect entanglement ($S_{CHSH} < 2\sqrt{2}$).*

The usual way to measure entanglement in the lab is with the $S_{CHSH}$ parameter. Assuming that the curves of coincidences as a function of $|\alpha-\beta|$ are sinusoidal and symmetric, the observed value of $S_{CHSH}$ is directly related with the standard visibility of the sinusoids:

$$S_{real} = 2\sqrt{2}.(N_{max}-N_{min})/(N_{max}+N_{min}) \qquad (7)$$

where $N_{max}$ ($N_{min}$) is the number of coincidences in the maximum (minimum) of the sinusoid. As the sinusoids for the different types of outcomes (even or odd) are assumed symmetric, then $N_{min}$ of the even-coincidence curves (i.e., at $|\alpha-\beta| = 90°$) is equal to $N_{min}$ of the odd-coincidence curves at $\alpha=\beta$. This, in turn, is only slightly different from the number of odd coincidences at $|\alpha-\beta|<<1$, for $\alpha=\beta$ is a minimum of the curve of odd coincidences. The contribution of odd elements in the series due to imperfect entanglement can be then estimated as $N_{min}$ in Eq.8. Defining $1- S_{real}/2\sqrt{2} \equiv \varepsilon$, then $N_{min} \approx \tfrac{1}{2}.m.\varepsilon$ ($\varepsilon<<1$). As in the case of accidental coincidences, the contribution caused by imperfect entanglement can be neglected only if the number of "true" odd elements $m.\varphi$ is much larger than $N_{min}$ or:

$$\varphi >> \tfrac{1}{2}.\varepsilon = \tfrac{1}{2}.(1- S_{real}/2\sqrt{2}) \qquad (8)$$

This bound is usually higher than the one in Eq.6.

*3.3 Imperfect efficiency (rate coincidences/singles < 1)*

Detectors have imperfect efficiency ($\eta<1$). Detection (or, better said, non-detection) is (supposedly) in-deterministic. This is another source of indeterminism that can mask the hypothesized smoothly decaying determinism. Experiments using cryogenic detectors [5,6] have reached total $\eta\approx0.78$.

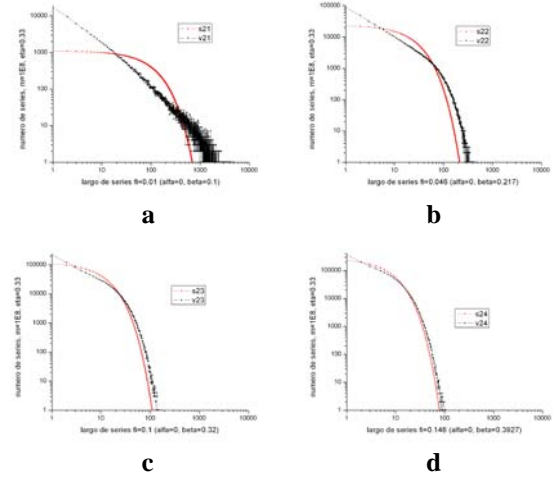

Figure 3: Numerically calculated distributions in the case $\eta=0.33$; (a) $\varphi= 0.01$, (b) $\varphi=0.046$, (c) $\varphi=0.1$, (d) $\varphi=0.146$. Note the distributions become indistinguishable for values of $\varphi$ smaller than in Fig.2 ($\varphi\approx0.1$ instead that $\varphi\approx0.5$).

In order to explore the influence of imperfect efficiency, the program code WQM is modified so that each detector "picks" the photons that are actually detected (and hence included in the calculus of coincidences), in an independent and random way. The rate of "picked" photons determines the total efficiency of the branch (say, the "+" output gate of the polarizer in station A). As an illustration, numerical results are plotted in Figure 3 for several values of $\varphi$ and assuming $\eta = 0.33$, the same for all detectors. For $\eta=1$, the deterministic and in-deterministic distributions can be easily distinguished for almost all values of $\varphi$ (Fig.2). For $\eta=0.33$, instead, they are practically indistinguishable already for $\varphi=0.1$ ($|\alpha-\beta|=18.4°$, Fig.3c). This is the consequence of the indeterminism introduced by the detection process. Numerical simulations show that the value of $\varphi$ that allows distinguishing the distributions decreases with $\eta$. There is, in consequence, a maximum useful value of $\varphi$ (say, $\varphi=0.1$ for $\eta=0.33$). In a real setup, efficiencies can be measured and included as parameters of the numerical simulation to estimate, in advance, whether that setup is suitable or not.

*3.4 An example.*

There is a window of values for φ that may allow observing the hypothesized slow decay of determinism. In practice, this window is limited from below by Eq.8, imperfect entanglement, and from above by imperfect efficiency. If the lower limit is higher than the upper one, then the hypothesized slow decay of determinism cannot be observed. In other words: not all values of {φ, η, $S_{real}$} allow detecting determinism, if it exists. As a rule of thumb (and not surprisingly) the experiment must have high entanglement and efficiency for that window to exist. The numbers are exigent, but attainable. F.ex., a set of parameters that allows distinguishing the distributions is: $S_{real} \geq 2.8$ and $\eta \geq 0.7$ (as in [7,8]), then the window is $5 \times 10^{-3} << \varphi \leq 0.25$, or $|\alpha-\beta|$ between 13º and 30º.

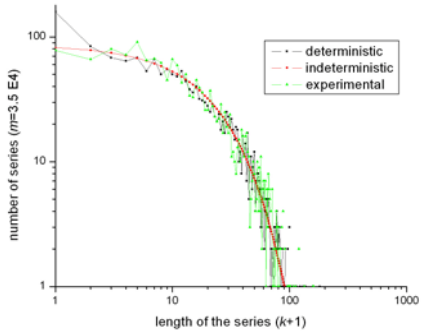

Figure 4: Experimental (green) and numerically calculated distributions (color codes as in Fig.1). The latter use the parameters corresponding to the former: $\eta = 0.1$, $\varphi = 0.048$, $S_{real} = 2.66$, $m$=35,632; $T_w = 2$ ns.

As an illustration of what is observed in practice, an actual distribution recorded in a time-resolved Bell's experiment (which was performed in 2022 with a different purpose [9]) is displayed in Figure 4. In this run $T_w$=2 ns and total number of coincidences is 35,632. It is plotted together with the distributions calculated according to WQM and $n(k)$ using the set of the experiments' measured parameters {φ,η, $S_{real}$} = {0.048, 0.1, 2.66}. According to the criteria discussed before, these values do not allow distinguishing determinism from indeterminism. For, from Eq.8: φ>>0.03, but for η=0.1 the numerical simulations indicate φ<0.01. The lower limit is higher than the upper one, so that there is no window. In fact, the three distributions are practically coincident, as expected.

**4. Conclusions.**

The apparently discontinuous change between determinism and indeterminism at some critical points challenges intuition and experience, and deserves to be studied. A classical deterministic and nonlocal model (WQM) suggests a smooth change to be possible.

It is shown that the boundary can be experimentally explored. Restrictions on the level of noise, of entanglement and of detectors' efficiency must be taken into account to open a window for a relevant observation. Outside this window, determinism, even if it exists, is blurred by indeterminism originated in the classical instruments. It is pertinent recalling that, according to the ideas in [1], there is no intrinsic indeterminism in QM. All indeterminism is eventually caused by the influence of the environment (to which, in general terms, classical instruments belong).

If the mentioned restrictions are fulfilled instead, the observation of deviations from $n(k)$ within that window (even if they do not fit the WQM distribution) would demonstrate not only that the boundary is continuous, but also that intrinsic quantum indeterminism does not exist. In addition to its importance from the foundational point of view, that observation would have a practical impact. It would demonstrate that there is no "certified randomness" in quantum random number generators, but just randomness by ignorance, as in classical physical generators [10].

**Appendix. About the WQM model.**

WQM assumes each particle in an entangled pair to carry the same, randomly varying, vector hidden variable **V**(t). The vector **V**(t) is transversal to the particles' direction of propagation; it is projected on each of the orthogonal axes of the polarizers. The modules of these projections are summed up in "memories" in each detector. When the sum in one memory becomes higher than a threshold value $u$, a particle is detected in the corresponding detector, and $u$ is subtracted from the memory. Therefore, although **V**(t) varies randomly, the mechanism of particle detection is deterministic. Knowing **V**(t) and the polarizer's orientation one knows, with certainty, when and where a particle is detected.

The model is nonlocal: if a particle is detected at station A, **V**(t) at B becomes "instantaneously" parallel to the axis of the polarizer's gate where the particle was detected. Leaving aside this nonlocal feature, the detection mechanism at B is identical to the one at A. If the nonlocal instruction is deleted, then WQM will predict the probability values of the semi-classical theory of radiation [11] (which are, of course, different from Eqs.1).

Note that, in the description above, detections in A influence detections in B, but not the other way. In the version of WQM used to plot the Figures in this paper, stations A and B switch roles after one particle detection. The series generated by the code are then analyzed to draw the distributions of strings (curves in black in the Figures).

WQM accurately reproduces the probabilities of detection predicted by QM even if η=1 and the angle settings are varied randomly. It is, in our knowledge, the simplest successful deterministic (but nonlocal) numerical simulation of Bell's experiment.

**Acknowledgments.**

Many thanks to Prof. K.Svozil for his observations to an earlier version of this manuscript. This work received support from the grants PUE 229-2018-0100018CO and PIP 00484-22, both from CONICET (Argentina).

**References.**